\documentclass[ 
,twocolumn 
,secnumarabic, superscriptaddress, nobibnotes, 
aps, prl]{revtex4}
\usepackage{amssymb}

\expandafter\ifx\csname package@font\endcsname\relax\else
 \expandafter\expandafter
 \expandafter
\expandafter
 \expandafter{\csname package@font\endcsname}
\fi
\newcommand{\bea}{\begin{eqnarray}}
\newcommand{\beq}{\begin{equation}}
\newcommand{\eea}{\end{eqnarray}}
\newcommand{\eeq}{\end{equation}}
\newcommand{\nn}{\nonumber}
\newcommand{\Frac}[2]{\frac{\displaystyle{#1}}{\displaystyle{#2}}}
\newcommand{\lsim}{\raise0.3ex\hbox{$\;<$\kern-0.75em\raise-1.1ex\hbox{$\=
sim\;$}}}
\newcommand{\gsim}{\raise0.3ex\hbox{$\;>$\kern-0.75em\raise-1.1ex\hbox{$\=
sim\;$}}}

\begin{document}

\title{Yukawa structure, flavour changing and CP violation in Supergravity.}
\author{G. G. Ross}
\affiliation{Dep. of Physics, Theoretical Physics, U. of Oxford, Oxford, 
OX1 3NP, UK}
\affiliation{Theory Division, CERN, CH-1211, Geneva  23, Switzerland}
\author{O. Vives}
\affiliation{Dep. of Physics, Theoretical Physics, U. of Oxford, Oxford, 
OX1 3NP, UK}

\begin{abstract}
The hierarchical structure of fermion masses and mixings strongly 
suggests
an underlying family symmetry. In supergravity any familon field
spontaneously breaking this symmetry necessarily acquires an F-term 
which
contributes to the soft trilinear couplings. We show, as a result, $\mu
\rightarrow e\gamma $ decay can receive large contributions from this 
source
at the level of current experimental bounds and thus this channel may
provide the first indication of supersymmetry and a clue to the 
structure of
the soft breaking sector. Using the mercury EDM\ bounds we find strong
bounds on the right handed down quark quark mixing angles that are
inconsistent with models relating them to neutrino mixing angles and 
favour
a near-symmetric form for the magnitude of the down quark mass matrix.
\end{abstract}

\maketitle


Our knowledge of the Supersymmetry breaking sector in the Minimal
Supersymmetric extension of the Standard Model (MSSM) is still very 
limited.
Only after the discovery of SUSY particles and the measurement of the
Supersymmetric spectrum we will be able to explore it in detail.
Nevertheless, we have already a lot of useful information on this sector
from experiments looking for indirect effects of SUSY particles in
low-energy experiments \cite{Annrev}. In fact, it was readily realized 
at
the beginning of the SUSY phenomenology era that large contributions to
Flavour Changing Neutral Currents (FCNC) and $CP$ violation phenomena 
were
expected in Supersymmetric theories with a generic soft breaking sector. 
The
absence of these effects much below the level most theorists considered
reasonable came to be known as the SUSY flavour and $CP$ problems. These
problems are closely related to the flavour structure of the soft 
breaking
terms and therefore to the source of flavour itself. The main solutions 
to
these problems are either universality of the soft breaking terms or
alignment with the Yukawa matrices.

In this paper we ask whether either of these solutions is sufficent in 
the
case the theory possesses a broken family symmetry which generates the
hierarchical structure of Yukawa couplings through a superpotential of 
the
form%
\begin{equation}
W=W^{hid}(\eta _{k})+\left( \frac{\displaystyle{\theta 
}}{\displaystyle{M}}%
\right) ^{\alpha _{ij}}H_{a}Q_{Li}q_{Rj}^{c}+\dots   \label{yukawas}
\end{equation}%
The hierarchy arises through effective operators ordered in terms of
powers of $(\theta /M)$ where $\theta $ is a familon field which
spontaneously breaks the family symmetry and $M$ is the mass of the 
mediator
transmitting the symmetry breaking to the quarks and leptons 
($M\leq
M_{Planck}$). We will show that neither universality of the soft 
breaking
terms nor alignment with the Yukawa matrices solutions can be exact and 
that
the soft breaking terms are necessarily nonuniversal and not aligned 
with
the Yukawa couplings. This is due to the fact that, in a theory with broken
local Supersymmetry, the familon field(s) necessarily acquires a
non-vanishing $F$-term \cite{abelservant}. This $F$-term then induces 
nonuniversal soft SUSY breaking $A$-terms which are not diagonalised 
when the fermion masses are diagonalised,
leading to FCNC and $CP$ violating processes.

We first prove the existence of a non-vanishing $F$ term. Consider the
supergravity lagrangian specified in terms of the superpotential $W$ and 
the
K\"{a}hler potential $K=\hat{K}(\eta ,\eta ^{\ast 
})+\sum_{i}K_{i}^{i}(\eta
,\eta ^{\ast })|\varphi _{i}|^{2}$ where $K$ is a real function of the
chiral superfields and $\eta $ are general fields belonging to the 
hidden
sector. The $F$-term contribution to the scalar potential is simply given
by,
\begin{equation}
V=e^{K}\left[ 
\sum_{i}(K^{-1})_{i}^{j}\tilde{F}^{i}\tilde{F}_{j}-3|W|^{2}%
\right]   \label{scalpot}
\end{equation}%
where $\tilde{F}^{i}=\partial W/\partial \phi _{i}+K_{i}^{i}\phi 
^{i\ast }W$
and $\tilde{F}_{i}=\partial W/\partial \phi ^{i\ast }+K_{i}^{i}\phi
_{i}W^{\ast }$ are related to the normalized supergravity $F$-terms by $%
F^{i}=e^{K/2}(K^{-1})_{i}^{i}\tilde{F}^{i}$ and $(K^{-1})_{j}^{i}$ is the 
inverse of the K\"{a}hler metric. In the
following, we require cancellation of the cosmological 
constant in the physical minimum, therefore from Eq.~(\ref{scalpot}), 
this implies that we do not need to consider the derivatives of $e^K$ in 
the minimization of the scalar potential. We study the case of broken 
supergravity in which $m_{3/2}=e^{K/2}\langle W\rangle \neq 0$. We will 
also have a non-vanishing
vev for a certain familon field $\theta $ after minimization of the 
scalar
potential. This field should have a small vev in units of the Plank mass 
to
generate the hierarchy in the Yukawa couplings which are proportional to
powers of $\theta $. The field $\theta$ 
belongs to a non-trivial
representation of the family group and is a singlet under the Standard 
Model
gauge group. We will also allow for further familon fields, $\chi _{i},$ 
transforming non-trivially under the family symmetry to acquire vevs. 
Given this, the form of the $\theta $ and $\chi $ dependence of the 
K\"ahler potential is fixed by the requirement that the
potential should be invariant under the family symmetry and the leading 
term is given by $K =\theta \theta^* +\sum_{i}\chi _{i} \chi^*_{i}$ 
where we have absorbed the constants
of proportionality in a redefinition of the fields. 
With this form of K\"ahler potential we have
$F^{\theta }=\langle \partial W/\partial \theta~+~W~\theta ^{\ast} \rangle$,
and the only way to avoid the conclusion that $F^{\theta }\geq 
m_{3/2}~\theta$ is if a cancellation occurs between the two terms 
\footnote{For general K\"ahler potentials depending on the hidden sector,
minimization of the potential with respect to the unnormalized 
$\theta$ is equal to minimization with respect to 
the rescaled $\theta$ up to higher order corrections in $\theta$}.

It is instructive  to
consider first the case that the $\theta $ dependent part of the
superpotential contributing to the potential does not involve the vev 
any other fields, giving $V(\theta )=\left| F^{\theta }\right| ^{2}-3|W|^{2}$,
and minimisation with respect to $\theta $ leads to
\begin{equation}
F^{\theta }\simeq - 3 m_{3/2}^{2}~\theta ~\left(\Frac{\partial 
^{2}W^*}{\partial \theta ^{*\,2}}\right)^{-1},  \label{fmin}
\end{equation}%
where we have set $\langle W\rangle=m_{3/2}$, used the condition
\begin{equation}
\langle\frac{\partial W}{\partial \theta }\rangle\simeq 
-m_{3/2}\langle\theta ^{\ast }\rangle,
\label{fsmall}
\end{equation}%
that is needed if $F^{\theta }$ is to be reduced below its natural value 
$\geq m_{3/2}\theta, $ and neglected $\theta~\partial W/ \partial \theta 
\simeq W ~\theta^2$ with respect to $W$, due to $\theta <<1$.
From Eq.~(\ref{fmin}) we see that the condition for an anomalously small 
$F^{\theta }$ is 
\bea
|\frac{\partial ^{2}W}{\partial \theta ^{2}}|>>m_{3/2}.
\label{smallcond}
\eea
It is now straightforward to check whether this possibility is realised 
for
various forms for $W(\theta ).$

We first consider the case that there is only one scale in the problem,
namely the Planck scale and that the fermion mass hierarchy is due to an
expansion in $\theta /M_{Planck.}$ \ One possibility, which has been 
widely
explored, is that there is an anomalous $U(1)$ family symmetry and only 
the
field $\theta $ acquires a vev close to the Planck scale, driven by the
requirement the anomalous $D-$term should be small. In this case the
superpotential cannot depend on $\theta $ except in combination with 
fields
which, for the moment, we require to have vanishing vevs, leading
immediately to the conclusion $F^{\theta }=\theta m_{3/2}$.

A second possibility is that the family symmetry is discrete and then one 
can have a dependence on $\theta $ of the form%
\begin{equation}
W=a+ (\theta /M_{Planck})^{p}  \label{monomial}
\end{equation}%
where we have allowed for a term, $a,$ coming from the hidden 
supersymmetry breaking sector and $p$ is the order of 
the discrete group. In this case 
$\partial ^{2}W/\partial \theta ^{2}\varpropto p(p-1)\theta
^{p-2}=(p-1)m_{3/2},$ where we have used Eq.~(\ref{fsmall}) to 
determine the
magnitude of $\theta .$ Inserted in Eq.~(\ref{fmin}) this suggests 
$F^{\theta
}$ is reduced by a factor $1/(p-1)$ compared to its natural value. 
However,
for $p>3,$ the true minimum of the potential following from Eq.~(\ref%
{monomial}) is at $\langle\theta \rangle=0.$
If $\theta $ has a renormalisable coupling $\theta XY$\ to other fields 
in
the theory, there are radiative corrections to the soft SUSY\ breaking 
terms
which must be included in the effective potential, namely the soft mass
squared and the soft trilinear term \cite{radiative}.  They can lead to 
the global minimum
having nonzero $\theta $ as desired. How do they affect 
our
conclusions? The soft mass only changes 
$\partial V/\partial \theta$ at ${\cal{O}}(m_{3/2}^2\theta )$ and so, 
in the large $p$ limit, does not affect the
derivation of Eq.~(\ref{fmin}). However the trilinear term contributes 
to $\partial V/\partial \theta$ at ${\cal{O}}((p-1)m_{3/2}\theta )$  
leading
to a shift in the position of the minimum at ${\cal{O}}(1).$ In turn this 
leads to ${\cal{O}}(\theta m_{3/2})$ corrections to $F^{\theta }$ in 
Eq.~(\ref{fmin}), i.e. $F^{\theta }$ is still of ${\cal{O}}(\theta m_{3/2})$.
This discusion can be repeated in the case there is a continuous 
symmetry with two charged fields obtaining vevs of the similar size along a 
$D$-flat direction with $W = a+ \chi^p \bar\chi^{q}$. In this case
$D$-flatness requires that
$Q \langle \chi\rangle  + \bar Q \langle \bar \chi \rangle =0$  with 
$Q$ and $\bar Q$ the charges under the continuous group. Using this 
relation we arrive again at the same result in this case.

To avoid this conclusion we must consider the case $\theta$ gets a mass 
in the Supersymmetric limit in order to satisfy Eq.~(\ref{smallcond})
\cite{rattazzi}. This requires the existence
of additional mass scales in the theory different from $M_{Plank}$ or 
$m_{3/2}$. We assume, as is 
expected in a string theory, that all mass 
scales in the theory are generated through spontaneous symmetry 
breaking and therefore we
consider the case there are additional
fields in the theory which acquire different vevs. This leads to the 
possibility that more than one term in the superpotential is important
\footnote{In the previous case one term dominates since $\theta /M_{Planck}$ 
is the only dimensionless parameter and is small.}. To demonstrate the 
possibilities
it is sufficient to consider just one extra field, $\phi ,$ and two 
terms in
the superpotential (setting the Planck scale to unity for clarity)
\begin{equation}
W(\theta ,\phi )=\theta ^{p}\phi ^{q}+\theta ^{p^{\prime }}\phi 
^{q^{\prime
}}  \label{poly}
\end{equation}%
Given that the field vevs are typically much greater than the electroweak
scale, we take $\phi ,$ like $\theta ,$ to be neutral under the SM group. The
form of the two terms in the superpotential must be determined by a symmetry
which may be the original family symmetry or may be a new one. The symmetry
can be either discrete, with $p$ and $p^{\prime }$ multiples of the order
parameter of the discrete group, or continuous if the fields acquires vevs
along a $D$-flat direction (in which case, to avoid $\langle \theta\rangle
\approx \langle \phi \rangle$ which would cause one of the terms in 
Eq.~(\ref{poly}) to
dominate, there must be additional\ non singlet fields which acquire vevs).
We include in the effective potential for $\phi $ a term $V_{\phi }$ which
forces it to acquires a vacuum expectation value. Replacing $\phi $ by its
vev Eq.~(\ref{poly}) becomes
\begin{equation}
W(\theta )=M_{1}\theta ^{p}+M_{2}\theta ^{p^{\prime }}.  \label{polyth}
\end{equation}%
with $M_{1}=\langle\phi \rangle^{q}/M_{Planck}^{p+q-3},$ 
$M_{2}=\langle\phi \rangle^{q^{\prime }}/M_{Planck}^{p^{\prime }+q^{\prime }
-3}$. The potential following from $W(\theta )$ has a non-trivial minima 
in the globally supersymmetric limit given by
\begin{equation}
\langle\theta _{0}\rangle^{p^{\prime }-p}=-pM_{1}/p^{\prime }M_{2}  \label{theta0}
\end{equation}%
For a minimum with $\langle\theta \rangle$ and $\langle\phi \rangle$ 
less than the Planck mass $(q-q^{\prime })$ and $(p^{\prime }-p)$ must 
be positive. The minimum occurs
through a cancellation between the two terms in $\partial W/\partial
\theta.$ One can readily show that, due to the relatively small effect of 
the last term in Eq.~(\ref{scalpot}), the global minimum of the full 
supergravity potential
can be close to this minimum. In this case, due to the cancellation 
between the two terms, the fact that $F^\theta \simeq 0$ given by  
Eq.~(\ref{fsmall}), does not relate the magnitudes of 
$\partial ^{2}W/\partial \theta ^{2}$ 
and $m_{3/2}$ as was done in the discrete case from Eq.~(\ref{monomial}). 
Therefore the $\theta $ mass is allowed to be much larger than $m_{3/2}$ 
and $F^{\theta }$ is suppressed. 
Solving for the minimum of the potential following from 
Eq.~(\ref{polyth})
one finds (in Planck units)
\begin{eqnarray*}
F^{\theta } &=&\left( \frac{3m_{3/2}}{W^{\prime \prime }(\theta _{0})}%
\right) m_{3/2}\theta _{0} \\
&\simeq& \frac{3m_{3/2}}{p^{\prime }(p^{\prime }-p)\left( 
\frac{p}{%
p^{\prime }}\right) ^{\frac{\left( p^{\prime }-2\right) }{(p^{\prime 
}-p)}%
}\langle\phi \rangle^{q^{\prime }+\frac{\left( p^{\prime }-2\right) \left( 
q-q^{\prime
}\right) }{(p^{\prime }-p)}}}~~ m_{3/2}\theta _{0}\nn
\end{eqnarray*}
The suppression depends on $\langle\phi \rangle.$ To bound $\langle\phi \rangle$ requires
minimisation of the full $V(\theta ,\phi )$ potential following from 
Eq.~(\ref%
{poly}) and $V_{\phi }.$ In the global limit%
\bea
V(\theta ,\phi )=\left| F^{\theta }\right| ^{2}+\left| F^{\phi 
}\right|
^{2}+V_{\phi }
\eea
If $\theta <\phi $ (requiring $(q-q^{\prime })/(p^{\prime }-p)>1)$ the
individual terms in $F^{\theta }$ and their derivatives  are greater 
than
those in $F^{\phi }.$ Thus minimisation with respect to $\theta $ will 
be approximately equivalent to minimisation of $\left| F^{\theta }\right| 
^{2}$
keeping $\phi $ fixed as was done following Eq.~(\ref{polyth}). Since $%
F^{\phi }$ breaks supersymmetry and contributes to scalar masses, the
solution to the hierarchy problem requires $\left| F^{\phi }\right| \leq
m_{3/2},$ bounding the allowed vevs. This, together with $\theta <\phi ,$
requires, up to a constant factor,
\bea
\phi ^{\frac{qp^{\prime }-pq^{\prime }}{p^{\prime }-p}-1}<m_{3/2}.
\label{bound}
\eea
Using Eq.~(\ref{theta0}), this bound allows $F^{\theta }$ to be
smaller than $m_{3/2}\theta _{0}$ provided 2$(q-q^{\prime })/(p^{\prime
}-p)>1,$ consistent with $\theta <\phi .$ To take a
simple example, the case $p=3,q=2,p^\prime =4,q^\prime =0$ gives 
$\theta
=3\phi ^{2}/4$, with 2$\theta ^{3}\phi <m_{3/2}.$ Thus $\theta
/M_{Planck}\leq 10^{-4}$ and $F^{\theta }\geq \sqrt{3}\left( \theta
/M_{Planck}\right) ^{3/2}m_{3/2}\theta _{0}$ allowing for a strong
suppression of $F^{\theta }.$

So, it is possible in the general case to suppress $F^{\theta }$
below $\langle \theta \rangle~ m_{3/2}.$ However this is not the case in 
many models. If the expansion parameter determining the fermion hierarchy is 
$\theta /M_{Planck}$, $q$ must be large ($q\geq 8$ for $\theta
/M_{Planck}\simeq 0.1).$ If the family symmetry is responsible for 
making $q$
large in Eq.~(\ref{poly}) there must be no fields acquiring vevs with 
family charge allowing lower order terms. Effectively this means that 
the family
structure will be determined by the field $\theta $ carrying a single sign
of family charge. Such models have been used to generate fermion mass
structure but they cannot reproduce the phenomenologically successful Gatto,
Sartori, Tonin (GST) relation which requires familon fields of both charges 
\cite{tasi}. Moreover the multiplet structure needed to get $q\geq 8$ 
looks very contrived.  In the case there are fields of both signs of
family charge, for example if $\chi $ and $\overline{\chi }$ acquire vevs
along the $D$-flat direction, we can generate the GST relation. To 
get a reduction in $%
F^{\chi }$ again requires further field(s), $\phi $. One can reproduce the
argument following from Eq.~(\ref{poly}) interpreting $\theta ^{p}=\sqrt{%
\chi\overline{\chi }}$. This time there must be
additional symmetries requiring $q\geq 8$ for $\chi /M_{Planck}\simeq 0.1$.
To avoid this conclusion requires the introduction of yet another scale $%
M<M_{Planck}$ to order the family hierarchy which complicates the theory
further, perhaps making it less believable. 

Given this, we consider it very likely that the familon field(s) will have
an $F$-term greater than or equal to its natural value and so we turn now to a
discussion of the phenomenological implications that follow if $F^{\theta } =
\beta \langle \theta \rangle m_{3/2}$ with 
$\beta ={\cal{O}}(1)$. In this case 
terms
involving $\theta $ contribute to the soft SUSY breaking terms and
these terms violate flavour conservation and $CP$ via the couplings in 
Eq.~(\ref{yukawas}). Using Eqs.~(\ref{yukawas},\ref{scalpot}) we can
determine the soft
breaking terms in the observable sector after SUSY breaking. This leads 
to
the trilinear terms \cite{soni},
\begin{equation}
A_{ij}\hat{Y}^{ij}=F^{\eta}\hat{K}_{\eta}Y^{ij}+\alpha_{ij}\
\frac{\displaystyle{e^{K/2}}}{\displaystyle{M}}\left( 
\frac{\displaystyle{%
\theta }}{\displaystyle{M}}\right) ^{\alpha _{ij}-1}\beta m_{3/2}\theta
\label{tri}
\end{equation}%
with $\hat{K}_{\eta}=\partial \hat{K}/\partial \eta$ and $%
Y^{ij}=e^{K/2}(\theta /M)^{\alpha _{ij}}$. The presence of $\alpha_{ij}$ 
in the right hand side is due to the dependence of the effective Yukawa
couplings on $\theta $. Note that the small parameter $\theta $
in $F^{\theta }$  does not affect the trilinear couplings because 
is
reabsorbed in the Yukawa coupling itself,
\bea
m_{3/2}\ \langle \theta \rangle \ \frac{\displaystyle{\partial 
Y^{ij}}}{%
\displaystyle{\partial \theta }}=\alpha _{i,j}\ m_{3/2}\ Y^{ij}.
\eea
From Eq.~(\ref{tri}) we see that, in any model which explains the hierarchy 
in
the Yukawa textures through nonrenormalizable operators, the trilinear
couplings are necessarily nonuniversal. In a similar way, we also
expect non-renormalizable contributions to the K\"{a}hler potential of 
the
kind $(\theta \theta ^{\ast }/M^{2})^{\alpha (i,j)}$. However these
contributions appear only at order $2\alpha (i,j)$ in $\theta /M$ with
respect to the dominant term $\mathcal{O}(1)$. So, in the following we
concentrate in the nonuniversal trilinear couplings.

Next we must check whether this breaking of universality does not 
contradict
any of the very stringent bounds from low energy phenomenology. Although we
do not have a complete theory of flavour \cite{flavourth} that provides 
the full field dependence of the low energy effective Yukawas, 
a fit to the fermion masses and mixing angles points to a definite texture 
for
the Yukawa matrices \cite{liliana},
\begin{equation}
\frac{M}{m_{3}}=\left(
\begin{array}{ccc}
0 & b\epsilon ^{3} & c\epsilon ^{3} \\
b^{\prime }\epsilon ^{3} & d\epsilon ^{2} & a\epsilon ^{2} \\
f\epsilon ^{m} & g\epsilon ^{n} & 1%
\end{array}%
\right) ,  \label{texture}
\end{equation}%
with $\epsilon _{d}=\sqrt{m_{s}/m_{b}}=0.15$ and $\epsilon_{u}=\sqrt{%
m_{c}/m_{t}}=0.05$ at the unification scale, and $a,b,b^{\prime},d,g,f$
coefficients ${\mathcal{O}}(1)$ and complex in principle. The coefficient $c$
is very sensitive to the magnitude of the elements below the diagonal 
and can be of ${\cal{O}}(1)$ or smaller. Our analysis will not depend on 
its value. In this texture the two undetermined elements, $(3,1)$ and 
$(3,2)$, determine
the unmeasured right-handed quark mixings. However, from the magnitude of the
eigenvalues we can constrain $m\geq 1$. In the context of a 
broken
flavour symmetry the hierarchy in the Yukawa matrices is generated by
different powers in the vevs of the familon fields, $\epsilon 
_{a}=\langle
\theta _{a}\rangle /M$. For the case of a single familon we can 
immediately calculate the nonuniversality in the trilinear terms, 
$(Y^{A})_{ij}\equiv
Y_{ij}A_{ij}$,
\begin{equation}
(Y^{A})_{ij}=A_{0}Y_{ij}+\beta m_{3/2}\ Y_{33}\left(
\begin{array}{ccc}
0 & 3b\epsilon ^{3} & 3c\epsilon ^{3} \\
3b^{\prime }\epsilon ^{3} & 2d\epsilon ^{2} & 2a\epsilon ^{2} \\
fm\epsilon ^{m} & gn\epsilon ^{n} & 0%
\end{array}%
\right)  \label{trilinear}
\end{equation}%
with $A_{0}=F^{\eta}\hat{K}_{\eta}$. For the case of several
familons the $F$-terms of different fields are expected to differ and 
although the coefficients will change from Eq.~(\ref{trilinear}) 
the proportionality of $Y^A$ and $Y$ will also be lost. 
Our results will not depend sensitively on variations of ${\cal{O}}(1)$ 
in these coefficients so we analyse the
particular case of Eq.~(\ref{trilinear}). The Yukawa texture in Eq.~(\ref%
{texture}) is diagonalized by superfield rotations in the so-called SCKM
basis, $\tilde{Y}=V_{L}^{\dagger }\cdot Y\cdot V_{R}$. However, in 
this
basis large off-diagonal terms necessarily remain in the trilinear
couplings, $\tilde{Y}^{A}=V_{L}^{\dagger }\cdot Y^{A}\cdot V_{R}$. The
phenomenologically relevant flavour off-diagonal entries in the basis of
diagonal Yukawa matrices are,
\begin{eqnarray}
(\tilde{Y}^{A})_{32} &\simeq &Y_{33}\beta \ m_{3/2}\ g\  
n \
\epsilon ^{n}+\dots ~~~~~~~~ \\
(\tilde{Y}^{A})_{21} &\simeq &Y_{33}\ \beta m_{3/2}\ \epsilon
^{3}(b^{\prime }+a\ (\frac{\displaystyle{b^{\prime 
}}}{\displaystyle{d}}\ g\
n\ \epsilon ^{n}-f\ m\ \epsilon ^{m-1}))  \nonumber
\end{eqnarray}%
The form of these matrices applies at the messenger scale which, being 
due
to supergravity, is $M_{Planck}$. Then we must use the MSSM 
Renormalization
Group Equations (RGE) \cite{RGE} to obtain the corresponding matrices at 
$M_W$. The main effect in this RGE evolution is a large 
flavour
universal gaugino contribution to the diagonal elements in the sfermion 
mass
matrices (see for instance Tables I and IV in \cite{wien}). In this 
minimal
supergravity scheme we take gaugino masses as 
$m_{1/2}=\sqrt{3}m_{3/2}$ and
sfermion masses $m_{0}^{2}=m_{3/2}^{2}$. So, the average masses at $M_W$ are,
\begin{eqnarray}
&m_{\tilde{q}}^{2}\simeq 6\cdot m_{1/2}^{2}+m_{0}^{2}\simeq 19\ 
m_{3/2}^{2}&
\nonumber \\
&m_{\tilde{l}}^{2}\simeq 1.5\cdot m_{1/2}^{2}+m_{0}^{2}\simeq 5.5\
m_{3/2}^{2}&  \label{average}
\end{eqnarray}%
The RG evolution of the trilinear terms is also similarly dominated by
gluino contributions and $3^{rd}$ generation Yukawa couplings. However, 
the
offdiagonal elements in the down and lepton trilinear matrices are 
basically
unchanged for $\tan \beta \leq 30$ \cite{wien}. From here we can obtain 
the
full trilinear couplings and compare with the experimental observables 
at
low energies. The so-called Mass Insertions (MI) formalism is very 
useful in this framework \cite{gabbiani}. 
The left--right MI are defined in the SCKM basis as 
$(\delta
_{LR})_{ij}=(m_{LR}^{2})_{ij}/m_{\tilde{f}}^{2}$, with 
$m_{\tilde{f}}^{2}$
the average sfermion mass.
\begin{table}[tbp]
\begin{center}
\begin{tabular}{||c|c|c|c|c||}
\hline\hline
$x$ & ${\sqrt{\left|\mbox{Im} \left(\delta^{d}_{LR} \right)_{12}^{2} 
\right|}
}$ & $\left|\left(\delta^{d}_{LR} 
\right)_{13}\right| $
& ${\left|\left(\delta^{l}_{LR} \right)_{12} \right|}$ & ${%
\left|\left(\delta^{l}_{LR} \right)_{23} \right|}$ \\ \hline
$0.3 $ & $1.1\times 10^{-5} $ & $1.3\times 10^{-2} $ & $6.9\times 
10^{-7} $
& $8.7\times 10^{-3} $ \\
$1.0 $ & $2.0\times 10^{-5} $ & $1.6\times 10^{-2} $ & $8.4\times 
10^{-7} $
& $1.0\times 10^{-2} $ \\
$4.0 $ & $6.3\times 10^{-5} $ & $3.0\times 10^{-2} $ & $1.9\times 
10^{-6} $
& $2.3\times 10^{-2} $ \\ \hline\hline
\end{tabular}%
\end{center}
\caption{MI bounds from $\protect\varepsilon ^{\prime 
}/\protect\varepsilon $%
, $b\rightarrow s\protect\gamma $, $\protect\mu \rightarrow 
e\protect\gamma $
and $\protect\tau \rightarrow \protect\mu \protect\gamma $ with 
$m_{\tilde q} = 500 \mbox{ GeV}$ and $m_{\tilde l} = 100 \mbox{ GeV}$ and 
different values of  $x=m_{\tilde{g}(\tilde \gamma)}^{2}/m_{\tilde{q}
(\tilde l)}^{2}$. These bounds scale
as $(m_{\tilde{f}}(\mbox{GeV})/500(100))^{2}$ for different average 
sfermion
masses.}
\label{tab:MI1}
\end{table}
We estimate the value of the $(2,1)$ LR MI as,
\begin{eqnarray}
&&\left( \delta _{LR}^{d}\right) _{21}\simeq \frac{\displaystyle{%
m_{b}\,\epsilon _{d}^{3}}}{\displaystyle{19\,m_{3/2}}}(b^{\prime 
}\,+a\,%
\frac{\displaystyle{b^{\prime }}}{\displaystyle{d}}\,g\,n\,\epsilon
_{d}^{n}-a\,f\,m\,\epsilon _{d}^{m-1})\simeq  \nonumber \\
&&(b^{\prime }\,+a\,\frac{\displaystyle{b^{\prime }}}{\displaystyle{d}}%
\,g\,n\,\epsilon _{d}^{n}-a\,f\,m\,\epsilon _{d}^{m-1})\,7.5\times 
10^{-6}
\label{eps'}
\end{eqnarray}%
using $m_{\tilde{q}}\simeq 500$ GeV corresponding to $m_{3/2}\simeq 120$ GeV
and $\epsilon _{d}\simeq 0.15$ \cite{liliana}. We can compare our 
estimate
for the mass insertion with the phenomenological bounds in Table 
\ref{tab:MI1} with 
$x=m_{\tilde{g}}^{2}/m_{\tilde{q}}^{2}\simeq 1$. Even
allowing a phase ${\mathcal{O}}(1)$,  necessary to contribute to 
$\varepsilon
^{\prime }/\varepsilon $, we can see the bound requires only $m\geq 1$ 
which
is already required to fit the fermion masses. Note however that in the 
presence of a phase, $\varepsilon ^{\prime }/\varepsilon $ naturally
receives a sizeable contribution from the $b^{\prime }$ term 
\cite{murayama}. 

Similarly, the MI corresponding to the $b \to s \gamma$ decay are, 
\begin{eqnarray}
\left(\delta^d_{LR}\right)_{3 2} \simeq \frac{\displaystyle{%
m_{3/2}\,m_b\,g\,n\, \epsilon_d^n}}{\displaystyle{19\,m_{3/2}^2}} \simeq 2.2
\times 10^{-3}\,g\,n\,\epsilon_d^n
\end{eqnarray}
again with $m_{3/2}\simeq 120$ GeV. This estimate is already of the same
order of the phenomenological bound for any $n$ and we do not get any new
constraint on $n$.

The situation is more interesting in the leptonic sector. Here, the 
photino contribution is dominant for LR mass insertions. In Table 
\ref{tab:MI1}, we show the rescaled bounds from Ref. \cite{gabbiani} for 
the present limits on the branching ratio. In this case, it seems reasonable 
to expect some kind of lepton-quark Yukawa unification. To generate the 
correct
muon and electron mass we follow Georgi and Jarlskog's suggestion and 
put a
relative factor of 3 in the $(2,2)$ entry. We also put a factor of $3$ 
in
the $(2,3)$ and $(3,2)$ entries as is required by non-Abelian models
which seek to explain the near equality in the down quark mass matrix of 
the $(2,2)$ and $(2,3)$ elements. With this we obtain, 
\begin{eqnarray}
&\left( \delta _{LR}^{e}\right) _{12}\simeq \frac{\displaystyle{m_{\tau
}\,\epsilon _{d}^{3}}}{\displaystyle{5.5\,m_{3/2}}}(b^{\prime 
}+\,9a\,\frac{%
\displaystyle{b^{\prime }}}{\displaystyle{d}}\,g\,n\,\epsilon
_{d}^{n}-3\,a\,f\,m\,\epsilon _{d}^{m-1})\simeq &  \nonumber \\
&(b^{\prime }\,+9\,a\,\frac{\displaystyle{b^{\prime 
}}}{\displaystyle{d}}%
\,g\,n\,\epsilon _{d}^{n}-3\,a\,f\,m\,\epsilon _{d}^{m-1})\,8.7\times
10^{-6}&  \label{mueg}
\end{eqnarray}%
where we take $m_{3/2}\simeq 120$ GeV corresponding to 
$m_{\tilde{l}}=280$
GeV. This result should be compared with the experimental bound from the
non-observation of $\mu \rightarrow e\gamma $ given by $\left( \delta
_{LR}^{e}\right) _{12}\leq 7\times 10^{-7}\ (280/100)^{2}=5.5\times 
10^{-6}.$ 
Note that in Eq.~(\ref{mueg}) there is an
unavoidable contribution from the $b^{\prime }$ entry, $\left( \delta
_{LR}^{e}\right) _{12}\simeq b^{\prime }\,8.7\times 10^{-6},$ exceeding
the experimental bound. To avoid this requires a
larger value of the slepton mass. For $m_{\tilde{l}}=320$ GeV 
our estimate would be just below the MI bound ($m_{3/2}=136$ GeV, 
$m_{\tilde{q}}=600$ GeV). In this conditions, the contributions of the 
second term in Eq.~(\ref{mueg}) does not lead to a new constraints 
on the value of $n$. However, the last term would be subdominant for 
$m \geq 2$ but $m=1$ would not be allowed up to $m_{\tilde{l}}=550$ GeV 
($m_{\tilde{q}}=1000$ GeV). To summarise, assuming a quark-lepton unification 
at $M_{GUT},$
nonuniversality in the trilinear terms predicts a large $\mu \rightarrow
e\gamma $ branching ratio even beyond the values expected from other 
sources such as SUSY seesaw \cite{SUSYseesaw}. This illustrates the point 
that $\mu \to e\gamma $ is a particularly sensitive probe of SUSY and 
the soft breaking sector.

Another interesting constraint is provided by Electric Dipole Moment 
(EDM)
bounds. Even in the most conservative case, where all soft SUSY breaking
parameters and $\mu $ are real, 
we know that the Yukawa matrices contain phases ${\mathcal{O}}(1)$. 
If the trilinear terms are nonuniversal, these
phases are not completely removed from the diagonal elements of $Y^{A}$ 
in the SCKM basis and hence can give rise to large EDMs \cite{abel,rattazzi}. 
However
the phase in the trilinear terms will be exactly zero at leading order 
in $\theta $ for any diagonal element \footnote{This leads to a weaker 
bound than that found in \cite{abel}.}. To see 
this we must take into account the fact that the eigenvalues, $D(\theta 
),$
and mixing matrices, $V_{L,R}(\theta )$ of the Yukawa matrix depend on 
$%
\theta $, $Y(\theta )=V_{L}(\theta )\,D(\theta 
)\,V_{R}(\theta )^{\dagger }$.%
The contribution to the trilinear terms is proportional to $\theta \,
\partial Y/\partial \theta .$ Evaluating this in the SCKM basis, we have,
\bea
V_{L}^{\dagger }\theta \frac{\displaystyle{\partial Y}}{\displaystyle{
\partial \theta }}%
V_{R}=V_{L}^{\dagger }\frac{\displaystyle{\theta \partial 
V_{L}}}{%
\displaystyle{\partial \theta }}D+\frac{\displaystyle{\theta \partial 
D}}{%
\displaystyle{\partial \theta }}+D\frac{\displaystyle{\theta \partial 
V_{R}^{\dagger }}%
}{\displaystyle{\partial \theta }}V_{R}
\eea
In this expression the dominant contribution in $\theta $ to a diagonal
element is controlled by the second term. This follows because $\theta \
\partial V/\partial \theta $ always adds at least a power of $\theta $ 
to
the diagonal element and therefore the first and third terms in the 
above
equation can only contribute to subdominant terms in the $\theta $ 
expansion
for the diagonal elements. The dominant second term is proportional to 
the
leading $\theta $ term in $Y_{ii}$ with a coefficient of proportionality
equal to its power in $\theta $ i.e. the phase is unchanged and is real 
in
the basis that the Yukawa couplings are real. Therefore any
observable phase in the diagonal elements will only appear at higher 
orders,
requiring $n\geq 1$ or $m\geq 2$ or through higher order contributions 
to
entries of the Yukawa matrix. Using this result, and assuming real $\mu 
$
and soft breaking terms, the EDMs have the form
\bea
\mbox{Im}\left( \delta _{LR}^{q,l}\right) _{11}\simeq 
\frac{\displaystyle{%
m_{1}}}{\displaystyle{R_{q,l}\,m_{3/2}}}\left( \epsilon 
^{n}\,n\,+\epsilon
^{m-1}\,(m-1)\,\right)
\eea
where we use 
$m_{1}=m_{3}\,\epsilon ^{4}bb^{\prime }/d$, we take
all unknown coefficients to be unity and assume
an ${\mathcal{O}}(1)$ phase which is observable in the basis of real masses. 
The coefficients
$R_{q}=19$ and $R_{l}=5.5$ take care of the RGE effects in the 
eigenvalues as before. So with $\epsilon _{d}=0.15$, $\epsilon _{u}=0.05$ 
and $m_{d}\simeq 10$ MeV, $m_{u}\simeq 5$ MeV and $m_{e}=0.5$ MeV we get,
\begin{eqnarray}
\mbox{Im}\left( \delta _{LR}^{d}\right) _{11} \simeq \left( \epsilon
_{d}^{n}\,n\,+\epsilon _{d}^{m-1}\,(m-1)\,\right) \,3.9\times 10^{-6}
\nonumber \\
\mbox{Im}\left( \delta _{LR}^{u}\right) _{11} \simeq \left( \epsilon
_{u}^{n}\,n\,+\epsilon _{u}^{m-1}\,(m-1)\,\right) \,1.9\times 10^{-6}
\nonumber \\
\mbox{Im}\left( \delta _{LR}^{e}\right) _{11} \simeq \left( \epsilon
_{d}^{n}\,n\,+\epsilon _{d}^{m-1}\,(m-1)\,\right) \,6.7\times 10^{-7}
\end{eqnarray}%
\begin{table}[tbp]
\begin{center}
\begin{tabular}{|c||c|c|c|c|}
\hline
$x$ & $\vert \mathrm{Im}(\delta_{11}^{d})_{LR}\vert$ & $\vert 
\mathrm{Im}%
(\delta_{11}^{u})_{LR} \vert$ & $\vert \mathrm{Im}(\delta_{22}^{d})_{LR}
\vert$ & $\vert \mathrm{Im}(\delta_{11}^{l})_{LR} \vert$ \\ \hline\hline
0.3 & $4.3\times 10^{-8}$ & $4.3\times 10^{-8}$ & $3.6\times 10^{-6}$ & 
$%
4.2\times 10^{-7}$ \\
1 & $8.0\times 10^{-8}$ & $8.0\times 10^{-8}$ & $6.7\times 10^{-6}$ & $%
5.1\times 10^{-7}$ \\
3 & $1.8\times 10^{-7}$ & $1.8\times 10^{-7}$ & $1.6\times 10^{-5}$ & $%
8.3\times 10^{-7}$ \\ \hline
\end{tabular}%
\end{center}
\caption{MI bounds from the mercury EDM for an average squark mass of 
$600
\mbox{ GeV}$ and for the electron EDM with an average slepton mass of 
$320
\mbox{ GeV}$ and different values of 
$x=m_{\tilde{g}}^{2}/m_{\tilde{q}}^{2}$%
. The bounds scale as 
$(m_{\tilde{q}(\tilde{l})}(\mbox{GeV})/600(320))$.}
\label{tab:MI2}
\end{table}
We compare these estimates with the phenomenological bounds in Table 
\ref{tab:MI2} \cite{hg,gabbiani}.  The bounds from the neutron and electron 
EDM do
not provide any new information on the structure of the Yukawa textures.
However, the mercury EDM bounds are much more restrictive and taking $%
x\simeq 1$ we find that, in the down sector, the case $n=1$, $m=2$ 
is not
allowed by EDM experiments and we require $n\geq 2$, $m\geq 3$. As we 
said
above, the same applies to subdominant corrections to $Y_{22}$ and
$Y_{12},Y_{21}$ where the first correction to the dominant terms can 
only be
$\epsilon ^{4}$ or $\epsilon ^{5}$ respectively to satisfy EDM bounds.

In summary, we have shown here that in a broken supergravity theory any
field that acquires a vev also has an $F$-term of order $\langle \theta
\rangle m_{3/2}$. These $F$-terms contribute unsuppressed to the trilinear
couplings and have observable effects in low energy phenomenology. The 
most
significant contribution, assuming a GUT type relation between the quark 
and
lepton masses, is to $\mu \rightarrow e\gamma .$ To keep this at the 
level
of current experimental bounds requires a slepton mass greater than 
$320~GeV$%
. At this level $\mu \rightarrow e\gamma $ should be seen by the 
proposed
experiments in the near future. There are also significant bounds coming
from the mercury EDM bounds which impose significant limits on the down
quark matrix elements below the diagonal responsible for right handed
mixing. These bounds disfavour large right handed mixing and thus 
disfavour $%
SU(5)$ based models in which the large (left handed) neutrino mixing 
angles
are related to large down quark right handed mixing angles. Of course 
these
strong bounds apply only to supergravity models with gravity as the
supersymmetry breaking messenger. Models with light messenger states, 
such
as gauge mediation models, have a much lower value for $m_{3/2}$ and 
this
determines the size of these flavour and $CP$ violating effects.

We are grateful for useful conversations with R. Rattazzi. We
acknowledge support from the RTN European project HPRN-CT-2000-0148. 
O.V. acknowledges partial support from the Spanish  MCYT FPA2002-00612
and DGEUI of the Gen. Valenciana grant GV01-94.

\end{document}